\documentclass[showpacs,twocolumn,pra]{revtex4}%
\usepackage{amsfonts}
\usepackage{amsmath}
\usepackage{amssymb}
\usepackage{graphicx}%
\providecommand{\U}[1]{\protect\rule{.1in}{.1in}}
\begin{document}
\title{Anderson Localization of cold atomic gases with effective spin-orbit
interaction in a quasiperiodic optical lattice}
\author{Lu Zhou$^{1,2}$, Han Pu$^{2}$ and Weiping Zhang$^{1}$}
\affiliation{$^{1}$Quantum Institute for Light and Atoms, Department of Physics, East China
Normal University, Shanghai 200062, China}
\affiliation{$^{2}$Department of Physics and Astronomy, and Rice Quantum Institute, Rice
University, Houston, TX 77251-1892, USA}

\begin{abstract}
We theoretically investigate the localization properties of a spin-orbit
coupled spin-1/2 particle moving in a one-dimensional quasiperiodic potential,
which can be experimentally implemented using cold atoms trapped in a
quasiperiodic optical lattice potential and external laser fields. We present
the phase diagram in the parameter space of the disorder strength and those
related to the spin-orbit coupling. The phase diagram is verified via
multifractal analysis of the atomic wavefunctions and the numerical simulation
of diffusion dynamics. We found that spin-orbit coupling can lead to the
spectra mixing (coexistence of extended and localized states) and the
appearance of mobility edges.

\end{abstract}

\maketitle

\section{Introduction}

Anderson localization (AL) is considered as a fundamental physical phenomenon,
which was first studied in the system of noninteracting electrons in a crystal
with impurities \cite{al}. AL predicts the absence of diffusion of electronic
spin, which stems from the disorder of crystal and is the result of quantum
interference. Since disorder is ubiquitous, AL is rather universal and can
occur in a variety of other physical systems including light waves
\cite{light-al} and atomic matter-waves
\cite{inguscio2008,aspect2008,aspect2007}. Due to its intimate relation with
metal-insulator transition, many interesting topics in AL such as the
interplay between nonlinearity and disorder
\cite{inguscio2010,zhou2011,modugno20092,modugno2011} are still under intense study.

In condensed-matter physics, spin-orbit (SO) coupling originates from the
interaction between the intrinsic spin of an electron and the magnetic field
induced by its movement. It connects the electronic spin to its orbital motion
and thus the electron transport becomes spin-dependent. SO interaction can
significantly affect AL and this problem had been addressed by a few works in
the electronic gas system \cite{hikami1980,kohmoto2008}.

The experimental realization of ultracold quantum gases, together with the
technique of optical lattice potential, have provided a powerful playground
for the simulation of condensed-matter systems. In this composite system, one
can achieve unprecedented control over almost all parameters by optical or
magnetical means. Specifically, pseudo-disorder can be generated by
superimposing two standing optical waves of incommensurate wavelengths
together. As a consequence, AL of atomic matter wave can take place, which had
been experimentally observed \cite{inguscio2008,inguscio2010,modugno2011} and
extensively studied in theory \cite{modugno2009,modugno20092}.

This work is motivated by the recent experimental realization of SO coupling
in ultracold atomic gas \cite{spielman2011,zwierlein2012,zhang2012,so-rmp}. We
investigate the impact of SO coupling on Anderson localization of a spin-1/2
particle using the system of cold atomic gases trapped in a quasiperiodic
one-dimensional (1D) optical lattice potential and simultaneously subject to
the laser-induced SO interaction. The similar topic had also been addressed in
\cite{edmonds2012,zhu2009}, with the focus on the localization properties of
relativistic Dirac particles with cold atoms in a light-induced gauge field.
Our model and method are different from theirs and we do not consider the
relativistic region.

The paper is organized as follows. Sec. \ref{sec_model} introduces the
theoretical model and tight-binding approximation is applied in Sec.
\ref{sec_t-b}. In\ Sec. \ref{sec_phase} we present the phase diagram and
discuss its implications. Sec. \ref{sec_multifractal} is devoted to the
multifractal analysis of the atomic wavefunction, from which the proposed
phase diagram is verified. The diffusion dynamics is studied in Sec.
\ref{sec_diffusion} for a initially localized Gaussian wavepacket. Finally we
conclude in Sec. \ref{sec_conclusion}.

\section{Model and Hamiltonian}

\label{sec_model}

\begin{figure}[tbh]
\includegraphics[width=8cm]{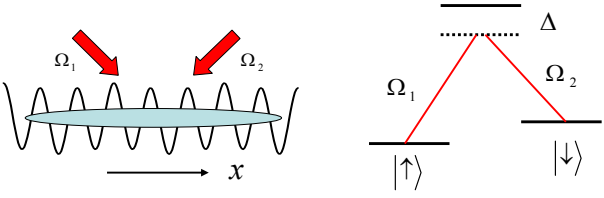}\caption{{\protect\footnotesize (Color
online) Schematic diagram showing the system under consideration.}}%
\label{scheme}%
\end{figure}

We consider the following model depicted in Fig. \ref{scheme}, cold atomic gas
with internal spin states $\left\vert \uparrow\right\rangle $ and $\left\vert
\downarrow\right\rangle $ confined in a spin-independent 1D quasiperiodic
optical lattice potential $s_{1}E_{R1}\sin^{2}\left(  k_{1}x\right)
+s_{2}E_{R1}\sin^{2}\left(  k_{2}x+\phi\right)  $ along the $x$-direction,
which is formed by combining two incommensurate optical lattice together
\cite{inguscio2008}. Here $k_{i}=2\pi/\lambda_{i}$ is the lattice wavenumber,
$s_{i}$ is the height of the lattice in unit of the recoil energy
$E_{R1}=h^{2}/2m\lambda_{1}^{2}$ and $\phi$ is the relative phase between the
two standing-wave modes (without loss of generality, we assume $\phi=0$ in the
following discussion). It is assumed that the depth of the lattice with
wavevector $k_{1}$ is deep enough to serve as the tight-binding primary
lattice. In the meanwhile, a pair of counter propagating Raman beams couple
the atomic states $\left\vert \uparrow,k_{x}=q\right\rangle $ and $\left\vert
\downarrow,k_{x}=q+2k_{R}\right\rangle $, which creats the effective SO
coupling \cite{spielman2011}.

In the basis composed of atomic pseudo-spin states $\left\{  \left\vert
\uparrow\right\rangle ,\left\vert \downarrow\right\rangle \right\}  $, the
single-particle Hamiltonian reads%
\begin{align*}
\hat{h}  &  =\left[  \frac{p_{x}^{2}}{2m}+s_{1}E_{R1}\sin^{2}\left(
k_{1}x\right)  +s_{2}E_{R1}\sin^{2}\left(  k_{2}x\right)  \right]  \hat{I}\\
&  +\frac{\Omega}{2}\left(
\begin{array}
[c]{cc}%
0 & e^{2ik_{R}x}\\
e^{-2ik_{R}x} & 0
\end{array}
\right)  ,
\end{align*}
with $\Omega$ the effective Raman coupling strength. Here we have assumed that
the Raman two-photon detuning is $0$. Then introduce the dressed pseudospin
states $\left\{  \left\vert \uparrow\right\rangle ^{\prime}=\left\vert
\uparrow\right\rangle e^{-ik_{R}x},\left\vert \downarrow\right\rangle
^{\prime}=\left\vert \downarrow\right\rangle e^{ik_{R}x}\right\}  $, which is
equivalent to performing a pseudo-spin rotation with the operator $\hat
{R}\left(  k_{R}\right)  =\exp\left(  -ik_{R}x\hat{\sigma}_{z}\right)  $, and
perform a global pseudo-spin rotation $\hat{\sigma}_{z}\rightarrow\hat{\sigma
}_{y}$, $\hat{\sigma}_{y}\rightarrow\hat{\sigma}_{x}$, $\hat{\sigma}%
_{x}\rightarrow\hat{\sigma}_{z}$ \cite{spielman2011,zwierlein2012}. In the new
basis, Hamiltonian $\hat{h}$ can then be rewritten as%
\begin{align}
\hat{h}  &  =\hat{h}_{SO}+V=\frac{\left(  p_{x}-\hat{A}\right)  ^{2}}%
{2m}+\frac{\Omega}{2}\hat{\sigma}_{z}+s_{1}E_{R1}\sin^{2}\left(  k_{1}x\right)
\nonumber\\
&  +s_{2}E_{R1}\sin^{2}\left(  k_{2}x\right)  , \label{s-p hamiltonian}%
\end{align}
in which $\hat{h}_{SO}$ describes single-particle motion\ in the presence of
the effective SO coupling, which is embodied in the vector potential $\hat
{A}=-m\lambda\hat{\sigma}_{y}$ $\left(  \lambda=\hbar k_{R}/m\text{
characterize SOC strength}\right)  $ and the effective Zeeman field $\Omega/2$.

$\hat{h}$ effectively describes a spin-$1/2$ particle moving in a 1D
quasiperiodic potential and subject to both the Zeeman field and equal
Rashba-Dresselhaus SO coupling, which can be used to simulate the
corresponding condensed matter system such as the motion of electron in one
dimentional semiconductor nanowire with disorder and SO interactions.

\section{Tight-binding approximation}

\label{sec_t-b}

In the language of quantum field theory, the total Hamiltonian describing our
system reads%
\begin{equation}
\hat{H}\left(  x\right)  =\int dx\hat{\Psi}^{\dagger}\left(  x\right)  \hat
{h}\left(  x\right)  \hat{\Psi}\left(  x\right)  , \label{hamilton}%
\end{equation}
with $\hat{\Psi}=\left(  \hat{\psi}_{\uparrow},\hat{\psi}_{\downarrow}\right)
^{T}$ the atomic field operators.

In the tight-binding limit, the atomic field operator can be expanded as
$\hat{\Psi}\left(  x\right)  =\sum_{j}w_{j}\left(  x\right)  \hat{c}_{j}$,
where $w_{j}\left(  x\right)  =w\left(  x-x_{j}\right)  $ is the Wannier state
of the primary lattice at the $j$-th site and $\hat{c}_{j}=\left(  \hat
{c}_{j\uparrow},\hat{c}_{j\downarrow}\right)  ^{T}$ are annihilation
operators. By considering that the tunneling takes place between nearest
neighbour sites and retaining only the onsite contribution of the secondary
lattice, one can achieve the following description of (\ref{hamilton}) (with
the energy measured in units of $E_{R1}$ and length scaled in units of
$k_{1}^{-1}$)%
\begin{widetext}
\begin{align*}
\hat{H}  &  =\sum_{j} \left[-\left(  \hat{c}_{j}^{\dagger}\hat{T}\hat{c}%
_{j+1}+H.c.\right)  -\Delta\cos\left(  2\pi\beta j\right)  \hat{c}%
_{j}^{\dagger}\hat{c}_{j}  +\frac{\Omega}{2}\hat{c}_{j}^{\dagger}\hat{\sigma}_{z}\hat{c}_{j} \right]\\
&  =\sum_{j} \left\{-J\left[  \hat{c}_{j}^{\dagger}\left(  \cos\pi\gamma-i\hat{\sigma
}_{y}\sin\pi\gamma\right)  \hat{c}_{j+1}+H.c.\right]  -\Delta\cos\left(  2\pi\beta j\right)  \hat{c}_{j}^{\dagger}\hat{c}%
_{j}+\frac{\Omega}{2}\hat{c}_{j}^{\dagger}\hat{\sigma}_{z}\hat{c}_{j} \right\} \,,
\end{align*}
where the tunneling amplitude $\hat{T}=J\exp\left(  -i/\hbar%
{\textstyle\int}
\hat{A}dl\right)  $ is obtained through Peierls subsititution
\cite{peierls,hofstadter1976}, which was also used in recent works to study
the impact of SO coupling on two-dimensional Bose-Hubbard model
\cite{cole2012,radic2012,Osterloh2005}. $%
{\textstyle\int}
\hat{A}dl=\hat{A}\left(  x_{j}-x_{j+1}\right)  =\hbar\pi\gamma\hat{\sigma}%
_{y}$ is the integral of the vector potential along the hopping path with
$\gamma=k_{R}/k_{1}$. $J$ is the tunneling amplitude without SO coupling,
$\beta=k_{2}/k_{1}$. $J$ and $\Delta$ can be calculated as%
\begin{align*}
J  &  =-\int dxw_{j+1}\left(  x\right)  \left[  -\frac{d^{2}}{dx^{2}}%
+s_{1}\sin^{2}x\right]  w_{j}\left(  x\right)  ,\\
\Delta &  =\frac{s_{2}\beta^{2}}{2}\int dx\cos\left(  2\beta x\right)
\left\vert w\left(  x\right)  \right\vert ^{2}.
\end{align*}

By writing $\left\vert \psi\right\rangle =\sum_{j,\sigma}\psi_{j,\sigma}%
\hat{c}_{j,\sigma}^{\dagger}\left\vert 0\right\rangle $, the stationary
Schr\"{o}dinger equation $\hat{H}\left\vert \psi\right\rangle =E\left\vert
\psi\right\rangle $ lead to%
\begin{subequations}
 \label{discrete eqns}%
\begin{gather}
-J\cos\left(  \pi\gamma\right)  \left(  \psi_{j+1,\uparrow}+\psi
_{j-1,\uparrow}\right)  -J\sin\left(  \pi\gamma\right)  \left(  \psi
_{j+1,\downarrow}-\psi_{j-1,\downarrow}\right) 
-\Delta\cos\left(  2\pi\beta j\right)  \psi_{j,\uparrow}+\frac{\Omega}{2}%
\psi_{j,\uparrow}=E\psi_{j,\uparrow} \,,\\
-J\cos\left(  \pi\gamma\right)  \left(  \psi_{j+1,\downarrow}+\psi
_{j-1,\downarrow}\right)  +J\sin\left(  \pi\gamma\right)  \left(
\psi_{j+1,\uparrow}-\psi_{j-1,\uparrow}\right) 
-\Delta\cos\left(  2\pi\beta j\right)  \psi_{j,\downarrow}-\frac{\Omega}%
{2}\psi_{j,\downarrow}=E\psi_{j,\downarrow} \,.
\end{gather}
\end{subequations}
\end{widetext}

The second terms on the left hand side of Eqs. (\ref{discrete eqns}), which are proportional to $J \sin(\pi \gamma)$, represent the spin-flipping tunneling which
arises from the effective SO interaction. In the absence of SO coupling $\left(
\gamma=0\text{, }\Omega=0\right)  $, spin-$\uparrow$ and $\downarrow$ components
are decoupled and we have%
\begin{equation}
-J\left(  \psi_{j+1}+\psi_{j-1}\right)  -\Delta\cos\left(  2\pi\beta j\right)
\psi_{j}=E\psi_{j}, \label{harper}%
\end{equation}
which represents the typical Harper equation \cite{harper} or the
Aubry-Andr\'{e} model \cite{aa}. Eq. (\ref{harper}) satisfies Aubry duality,
as can be seen by performing the transformation $\psi_{j}=\sum_{m}\tilde{\psi
}_{m}e^{im\left(  2\pi\beta j\right)  }$, insert it into Eq. (\ref{harper})
will lead to%
\begin{equation}
-\frac{\Delta}{2}\left(  \tilde{\psi}_{m+1}+\tilde{\psi}_{m-1}\right)
-2J\cos\left(  2\pi\beta m\right)  \tilde{\psi}_{m}=E\tilde{\psi}_{m},
\label{harper1}%
\end{equation}
Eqs. (\ref{harper}) and (\ref{harper1}) are identical at $\Delta/J=2$. Since
the transformation made above represents the typical Fourier transform which
transforms localized states to extended states and vice versa, then the
critical point $\Delta/J=2$ is identified as the transition point between the
localized states and extended states, i.e., all the single-particle states are
extended when $\Delta/J<2$ and localized when $\Delta/J>2$.

The properties of the Aubry-Andr\'{e} model, as represented by Eq.~(\ref{harper}), have been theoretically studied 
\cite{kohmoto1983,thouless1983} and it can be implemented in systems of
Bloch electrons \cite{hofstadter1976} and cold atoms \cite{inguscio2008}. We
have also studied this model with $\Delta$ dressed by a cavity field through
nonlinear feedback \cite{zhou2011}.

A similar transformation $\psi_{j,\sigma}=\epsilon_{\sigma}\sum_{m}\tilde
{\psi}_{m,\sigma}e^{im\left(  2\pi\beta j\right)  }$ $\left(  \epsilon
_{\uparrow}=1,\epsilon_{\downarrow}=i\right)  $ made to Eqs.
(\ref{discrete eqns}) will lead to%
\begin{widetext}
\begin{subequations}
\label{discrete eqns_t}%
\begin{gather}
-\frac{\Delta}{2}\left(  \tilde{\psi}_{m+1,\uparrow}+\tilde{\psi
}_{m-1,\uparrow}\right)  +2J\sin\left(  \pi\gamma\right)  \sin\left(
2\pi\beta m\right)  \tilde{\psi}_{m,\downarrow}
-2J\cos\left(  \pi\gamma\right)  \cos\left(  2\pi\beta m\right)  \tilde{\psi
}_{m,\uparrow}+\frac{\Omega}{2}\tilde{\psi}_{m,\uparrow}=E\tilde{\psi
}_{m,\uparrow} \,,\\
-\frac{\Delta}{2}\left(  \tilde{\psi}_{m+1,\downarrow}+\tilde{\psi
}_{m-1,\downarrow}\right)  +2J\sin\left(  \pi\gamma\right)  \sin\left(
2\pi\beta m\right)  \tilde{\psi}_{m,\uparrow}
-2J\cos\left(  \pi\gamma\right)  \cos\left(  2\pi\beta m\right)  \tilde{\psi
}_{m,\downarrow}-\frac{\Omega}{2}\tilde{\psi}_{m,\downarrow}=E\tilde{\psi
}_{m,\downarrow}\,. 
\end{gather}
\end{subequations}
\end{widetext}
A comparison between Eqs.~(\ref{discrete eqns}) and Eqs.~(\ref{discrete eqns_t}) shows that the presence of the spin-flipping tunneling terms breaks the duality. This distinguishes our current work from that reported in Ref.~\cite{kohmoto2008}, in which the authors studied a system of two-dimensional (2D) electrons on a square lattice subject to Rashba spin-orbit coupling and immersed in a perpendicular
uniform magnetic field. In this system, it has been shown \cite{kohmoto2008} that a gneralized Aubry duality is preserved when tunneling along the two perpendicular lattice directions are exchanged. Such an operation is not available in our system as ours is an intrinsically 1D model.

Due to the lack of duality in the current model, it
is not clear whether there exists a phase transition between the localized and
extended states in the present system. In addition, what effect will SO
coupling take? Will it enhance the tendency to localization or delocalization?
We focus on these problems in the following discussion.

\section{Phase diagram analysis}

\label{sec_phase}

Here we follow the method in \cite{kohmoto2008} to map the phase diagram using
a quantity called the total width of all the energy bands $B$, which had been
proved to be useful in investigating phase transition in a quasiperiodic
system \cite{kohmoto2008,thouless1983,kohmoto1983}. In order to observe its
property, as people usually do, we first choose the optical lattice wavelength
ratio $\beta$ to be $\beta_{n}=F_{n}/F_{n+1}=p/q$, in which $F_{n}$ is the
$n$-th Fibonacci number defined by the recursion relation $F_{n+1}%
=F_{n}+F_{n-1}$ with $F_{0}=F_{1}=1$. When $n\rightarrow\infty$, $\beta
_{n}\rightarrow\left(  \sqrt{5}-1\right)  /2$, which is the inverse of the
golden ratio.

Since $p$ and $q$ are integers prime to each other, the system is periodic
with the period $q$. Under the periodic boundary condition, according to
Bloch's theorem, $\psi_{i+q,\sigma}=e^{ik_{x}q}\psi_{i,\sigma}$. Eqs.
(\ref{discrete eqns}) then reduce to an eigenvalue problem $\mathcal{H}%
\psi=E\psi$ with $\psi=\left(  \psi_{1,\uparrow},\psi_{1,\downarrow}%
,\psi_{2,\uparrow},\psi_{2,\downarrow},\cdot\cdot\cdot,\psi_{q-1,\uparrow
},\psi_{q-1,\downarrow},\psi_{q,\uparrow},\psi_{q,\downarrow}\right)  $ and
the $2q\times2q$ matrix $\mathcal{H}$ takes the following form%

\[
\mathcal{H}=\left(
\begin{array}
[c]{ccccccc}%
H_{1} & L & 0 & \cdots &  & 0 & e^{-ik_{x}q}L^{\dagger}\\
L^{\dagger} & H_{2} & L & 0 &  &  & 0\\
0 & L^{\dagger} & H_{3} & L & 0 &  & \\
& 0 & \cdots &  &  &  & \\
&  &  &  & \cdots & L & 0\\
0 &  &  & 0 & L^{\dagger} & H_{q-1} & L\\
e^{ik_{x}q}L & 0 &  &  & 0 & L^{\dagger} & H_{q}%
\end{array}
\right)
\]
with%
\[
H_{j}=\left(
\begin{array}
[c]{cc}%
-\Delta\cos\left(  2\pi\beta j+\phi\right)  +\frac{\Omega}{2} & 0\\
0 & -\Delta\cos\left(  2\pi\beta j+\phi\right)  -\frac{\Omega}{2}%
\end{array}
\right)
\]
and $L=\left(
\begin{array}
[c]{cc}%
-J\cos\left(  \pi\gamma\right)  & -J\sin\left(  \pi\gamma\right) \\
J\sin\left(  \pi\gamma\right)  & -J\cos\left(  \pi\gamma\right)
\end{array}
\right)  $. The Hermite matrix $\mathcal{H}$ can be diagonalized with $2q$
real eigenvalues $E_{i}\left(  k_{x}\right)  $, which form $2q$ energy bands
as the function of $k_{x}$ in the first Brillouin zone $q\left\vert
k_{x}\right\vert \leq\pi$.

In the absence of SO coupling, the energy bands are degenerate for
spin-$\uparrow$ and $\downarrow$, with the band edges locate at $k_{x}=0$ and
$k_{x}=\pm\pi/q$, so $B$ can be calculated by $B=%
{\textstyle\sum\nolimits_{i=1}^{2q}}
\left\vert E_{i}\left(  0\right)  -E_{i}\left(  \pm\pi/q\right)  \right\vert
$. It was first demonstrated in \cite{kohmoto1983} that for extended states
with $\Delta/J<2$, $B$ approaches a finite value for $q\rightarrow\infty$;
while for localized states $\Delta/J>2$, $B$ rapidly tends to zero as
$q\rightarrow\infty$; at the critical point $\Delta/J=2$, $B\approx q^{-1}$.
In this manner, the critical value $\Delta=\Delta_{c}$ signaling AL transition
can be determined by observing the property of $B$ as a function of the period
$q$.

\begin{figure}[tbh]
\includegraphics[width=8cm]{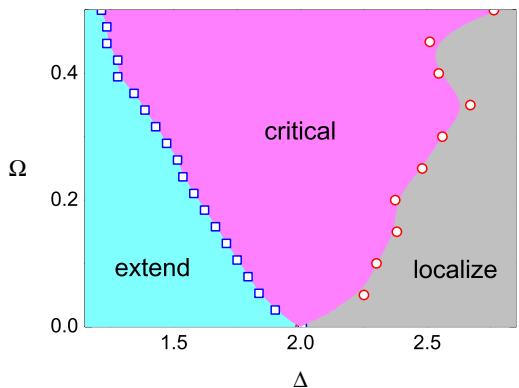}\caption{{\protect\footnotesize (Color
online) Phase diagram for Anderson localization of SO coupled BEC in 1D
quasiperiodic lattice in the parameter space }$\Delta$%
{\protect\footnotesize -}$\Omega${\protect\footnotesize . }$\gamma
=0.7${\protect\footnotesize , }$\Delta$ {\protect\footnotesize and }$\Omega$
{\protect\footnotesize are estimated in units of }$J${\protect\footnotesize .}%
}%
\label{fig_phase_d-o}%
\end{figure}

\begin{figure}[tbh]
\includegraphics[width=8cm]{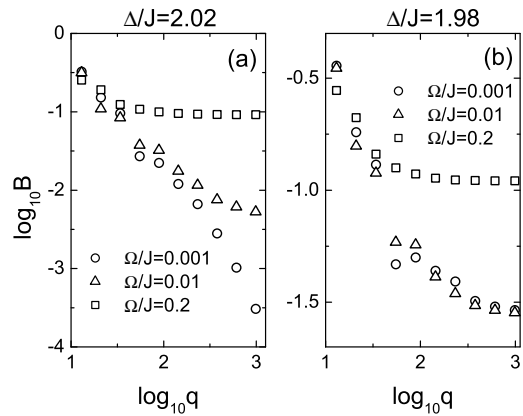}\caption{{\protect\footnotesize Total bandwidth
versus period }$q${\protect\footnotesize . At }$q\rightarrow\infty$
{\protect\footnotesize the system becomes quasiperiodic. The parameters are
specified in the figure.}}%
\label{fig_bw}%
\end{figure}

Now taking SO coupling into account, we anticipate that the phase diagram is
composed of three different phases: (i) Extended phase in which the energy
spectrum is purely continuous and all the eigenstates are extended; (ii)
Localized phase is characterized by purely dense point and all the
wavefunctions are localized; (iii) The energy spectrum is mixed in the
critical phase with extended eigenstates coexist with localized ones. Using
the method described above, $\Delta_{c}$ is determined\ as a function of
$\Omega$, which is indicated in the phase diagram of Fig. \ref{fig_phase_d-o}
by the line separating the regions signaling localized phase and critical
phase. Calculation is performed with the parameter of $\gamma=0.7$, which is
used throughout the paper and can be experimentally realized for $^{87}$Rb
atoms by adjusting the angle between Raman beams \cite{spielman2011}. At
$\Omega=0$, AL transition occurs at $\Delta/J=2$, reminiscent of the situation
without SO coupling. This can be understood from Eq. (\ref{s-p hamiltonian})
by that SO interaction can be removed from the Hamiltonian through a unitary
transformation with the operator $\hat{S}=\exp\left(  -ix\hat{\sigma}_{y}%
/2\xi\right)  $ when $\Omega=0$. Examples of data are shown in Fig.
\ref{fig_bw}(a). At $\Delta/J=2.02$, $B$ tends to zero for $\Omega
<\Omega\left(  \Delta_{c}\right)  $ and $B$ tends to a finite value for
$\Omega>\Omega\left(  \Delta_{c}\right)  $.

Due to that duality is broken by the SO coupling here, the boundary between
extended phase and critical phase is not related to that separates localized
phase and critical phase, which is different from \cite{kohmoto2008}. The
extended phase and critical phase cannot be differentiated by examing the
properties of $B$, since the energy spectrum contain absolutely continuous
parts in both these two phases, which leads $B$ to a finite value in the
quasiperiodic limit, as shown in Fig. \ref{fig_bw}(b).

The localization property of the atomic wavefunction can be characterized with
the inverse participation ratio (IPR) which is defined as%
\[
P^{-1}=\sum_{j=1}^{N}\left(  \left\vert \psi_{j,\uparrow}\right\vert
^{4}+\left\vert \psi_{j,\downarrow}\right\vert ^{4}\right)  ,
\]
in which $N$ is the number of lattice sites, $\psi_{j,\uparrow\left(
\downarrow\right)  }$ are solutions of Eqs. (\ref{discrete eqns}) and fulfil
the normalization condition $%
{\textstyle\sum\nolimits_{j}}
\left(  \left\vert \psi_{j,\uparrow}\right\vert ^{2}+\left\vert \psi
_{j,\downarrow}\right\vert ^{2}\right)  =1$. IPR reflects the inverse of the
number of the lattice sites being occupied by the atoms. For extended states,
$P^{-1}\rightarrow1/N$ and approach $0$ for large $N$. While for localized
states, IPR tends to a finite value and a larger value of $P^{-1}$ means that
the atoms are more localized in space.

\begin{figure}[tbh]
\includegraphics[width=8cm]{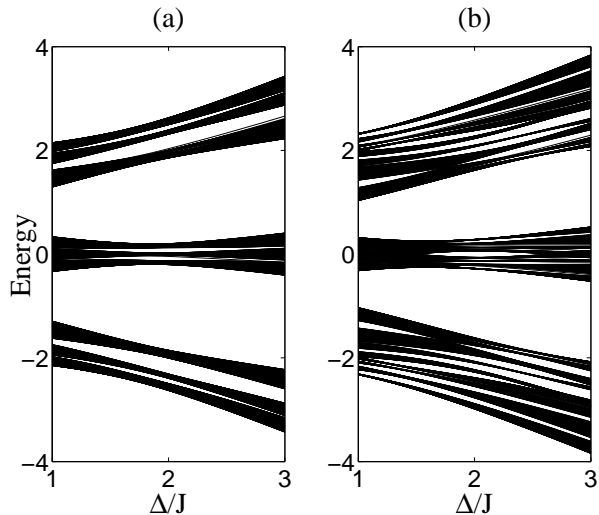}
\caption{{\protect\footnotesize Eigenenergies obtained from
numerical diagonalization of Hamiltonian matrix, as a function of
}$\Delta /J${\protect\footnotesize . Calculations are performed
under periodic
condition with }$\beta=610/987${\protect\footnotesize , }$N=987$%
{\protect\footnotesize , }$\gamma=0.7$ {\protect\footnotesize and (a) }%
$\Omega=0.1${\protect\footnotesize ; (b) }$\Omega=1${\protect\footnotesize .}}%
\label{fig_energy spectrum} \end{figure}

We use IPR to determine the boundary separating the extended phase and
critical phase, which is identified by the turning point of IPR as a function
of $\Delta/J$, as those had been done in many previous works
\cite{dufour2012,sarma2010,zhou2011,modugno2009}. The calculation is performed
with $\beta=F_{14}/F_{15}=610/987$ and $N=F_{15}=987$ under periodic boundary
condition. This give the extend-critical phase boundary shown in Fig.
\ref{fig_phase_d-o}, which indicate that with the increase of the Rabi
frequency $\Omega$ the system is more likely to start become localized. In
order to understand this, we plot the energy spectrum as a function of
$\Delta/J$. When $\Omega$ is relatively small, the spectrum shown in Fig.
\ref{fig_energy spectrum}(a) possesses similar properties as that in the
absence of SO coupling: Along with the increase of $\Delta/J$, two major gaps
devide the spectrum into three parts, each of which in turn devides into three
smaller parts, and so on. This is because the value of $1/\beta$ lies between
$2$ and $3$. The spectrum of localized states is then characterized by the
presence of an infinite number of gaps and bands. The effective Zeeman term
which is propotional to $\Omega$, in combination with SO coupling and the
lattice structure, take the effect of opening gaps between different energy
bands, as shown in Fig. \ref{fig_energy spectrum}(b). So the critical value
$\Delta_{c}$ takes a smaller value with the increase of $\Omega$.

\section{Multifractal analysis of wavefunctions}

\label{sec_multifractal}

To check the proposed phase diagram, we investigate the scaling property of
the wavefunctions using the method of multrifractal analysis described in
\cite{kohmoto2008}. Take the period of the lattice to be Fibonacci number
$F_{n}$, from the wavefunctions $\left\{  \psi_{j,\sigma}\right\}  $ we have
the probability $p_{j}=\left\vert \psi_{j,\uparrow}\right\vert ^{2}+\left\vert
\psi_{j,\downarrow}\right\vert ^{2}$, which is normalized as $%
{\textstyle\sum\nolimits_{j=1}^{F_{n}}}
p_{j}=1$. The scaling index $\alpha$ for $p_{j}$ is defined as $p_{j}%
=F_{n}^{-\alpha}$. We then assume that the number of sites satisfying the
scaling is propotional to $F_{n}^{f_{n}\left(  \alpha\right)  }$, $f\left(
\alpha\right)  $ can be calculated as $f\left(  \alpha\right)  =\lim
_{n\rightarrow\infty}f_{n}\left(  \alpha\right)  $.

The localization properties of wavefunctions are characterized by $f\left(
\alpha\right)  $ in the following manner: For extended wavefunctions, all the
lattice satisfy $p_{j}\sim F_{n}^{-1}$, so $f\left(  \alpha\right)  $ is fixed
at $f\left(  \alpha=1\right)  =1$. On the other hand, a localized wavefunction
has a nonvanishing probability only on a finite number of sites. These sites
have $\alpha=0$ $\left[  f\left(  0\right)  =0\right]  $\ and other sites with
probability zero have $\alpha=\infty$ $\left[  f\left(  \infty\right)
=1\right]  $. For critical wavefunctions, $\alpha$ has a distribution, which
means that $f\left(  \alpha\right)  $ is a smooth function defined on a finite
interval $\left[  \alpha_{\min,}\alpha_{\max}\right]  $.

Numerically we calculate $f_{n}\left(  \alpha\right)  $ for Fibonacci indices
$n$ and extrapolate them to $n\rightarrow\infty$. One can then discriminate
extended, localized and critical wavefunction by examing the minimum value of
$\alpha$ in the following manner%
\begin{align*}
\text{extended wavefunction }\alpha_{\min}  &  =1\text{, }\\
\text{critical wavefunction }\alpha_{\min}  &  \neq0,1\text{, }\\
\text{localized wavefunction }\alpha_{\min}  &  =0\text{.}%
\end{align*}

\begin{figure}[tbh]
\includegraphics[width=8cm]{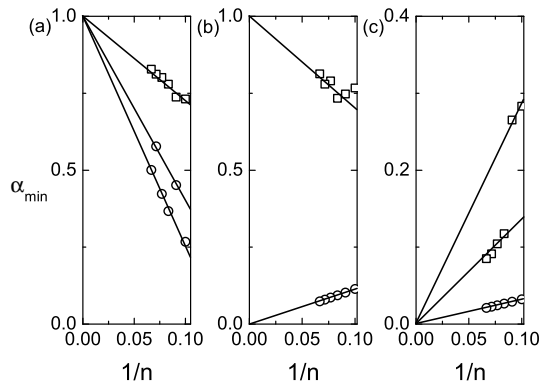}\caption{$\alpha_{\min}$
{\protect\footnotesize versus }$1/n$ {\protect\footnotesize for the
wavefunctions of the lowest band }$\bigcirc$ $\left(  i=1\right)  $
{\protect\footnotesize and the centre band }$\square$ $\left(  i=F_{n}\right)
${\protect\footnotesize . (a) }$\Omega/J=0.2${\protect\footnotesize , }%
$\Delta/J=1.5${\protect\footnotesize ; (b) }$\Omega/J=0.4$%
{\protect\footnotesize , }$\Delta/J=1.5${\protect\footnotesize ; (c) }%
$\Omega/J=2.5${\protect\footnotesize , }$\Delta/J=0.1$
{\protect\footnotesize corresponds to extend phase, critical phase and
localize phase, respectively.}}%
\label{fig_amin}%
\end{figure}$\alpha_{\min}$ is calculated for example wavefunctions and the
results are shown in Fig. \ref{fig_amin}. The wavefunction of the lowest band
is denoted by $i=1$, while that at the centre of the energy spectrum is
denoted by $i=F_{n}$. First, for $\Omega/J=0.2$, $\Delta/J=1.5$ at which the
system is in the extended phase according to the phase diagram Fig.
\ref{fig_phase_d-o}, $\alpha_{\min}$ extrapolates to $1$ for both $i=1$ and
$i=F_{n}$, as shown in Fig. \ref{fig_amin}(a), indicating that the energy
spectrum is purely continuous and all the wavefunctions are extended. The
point $\Omega/J=0.4$, $\Delta/J=1.5$ corresponds to the critical phase in Fig.
\ref{fig_phase_d-o}, and in Fig. \ref{fig_amin}(b) one can find out that
$\alpha_{\min}$ extrapolates to $0$ for $i=1$ and $\alpha_{\min}$ extrapolates
to $1$ for $i=F_{n}$. This suggests that the wavefunction at the edge of the
energy spectrum is localized while that at the centre is extended, which
indicates the existance of mobility edges.

The appearance of mobility edges here can be understood as the result of the
breaking of original self-duality via SO interaction, which can also be
aroused by other effects such as hopping beyond neighbouring lattice sites
\cite{boers2007,sarma2010}. We would like to note that, even if the duality is
preserved, SO coupling can also lead to the appearance of critical phase and
mobility edges, as those had been demonstrated in \cite{kohmoto2008}.

\section{Diffusion dynamics}

\label{sec_diffusion}

In realistic experiment, localization properties can be investigated by
loading the SO-coupled BEC into the quasiperiodic potential and observing its
transportation along the lattice \cite{inguscio2008}. The equations-of-motion
associated with Eqns. (\ref{discrete eqns}) are%
\begin{align}
i\frac{\partial\Psi_{j}}{\partial t}  &  =-Je^{-i\pi\gamma\hat{\sigma}_{y}%
}\left(  \Psi_{j+1}+\Psi_{j-1}\right) \nonumber\\
&  +\left[  -\Delta\cos\left(  2\pi\beta j\right)  +\frac{\Omega}{2}%
\hat{\sigma}_{z}\right]  \Psi_{j}, \label{eq_motion}%
\end{align}
where $\Psi_{j}=\left(  \psi_{j,\uparrow},\psi_{j,\downarrow}\right)  ^{T}$.
We study the diffusion of ultracold atomic gas in quasiperiodic optical
lattice with Eq. (\ref{eq_motion}) by taking the initial atomic wavefunction
to be a localized Gaussian wavepacket with width $a$
\[
\Psi_{j}\left(  t=0\right)  =\left(  2a\sqrt{\pi}\right)  ^{-1/2}%
e^{-j^{2}/2a^{2}}\binom{1}{i},
\]
in which we take $a=5$ in the following calculation.\ Here we assume that the
atomic wave packets initially lie in the centre at $j=0$ with the system size
of 2000 lattice sites. The numerical simulation is performed with vanishing
boundary condition and during the time evolution the atomic wavepacket never
reaches the boundaries so that the effect of boundary condition does not appear.

To measure the localization, we use the width of the wave-packet defined as%
\[
w=\sqrt{\left\langle \left(  \Delta x\right)  ^{2}\right\rangle }=\left\{
\sum_{j}j^{2}\left(  \left\vert \psi_{j,\uparrow}\right\vert ^{2}+\left\vert
\psi_{j,\downarrow}\right\vert ^{2}\right)  \right\}  ^{1/2}.
\]
In the absence of SO coupling, the time evolution of $w\left(  t\right)  $ can
be parametrized as $w\left(  t\right)  \sim t^{\eta}$
\cite{modugno20092,hiramoto1988}, and its property is intimately related to
the localization properties of the system:

(i) in the extended phase of $\Delta/J<2$, the wavepacket will experience
ballistic expansion with $\eta=1$;

(ii) at the critical point of $\Delta/J=2$ the wavepacket subject to
subdiffusion with $\eta\sim0.5$;

(iii) the wavepacket is localized when $\Delta/J>2$ and $\eta=0$%
.\begin{figure}[tbh]
\includegraphics[width=8cm]{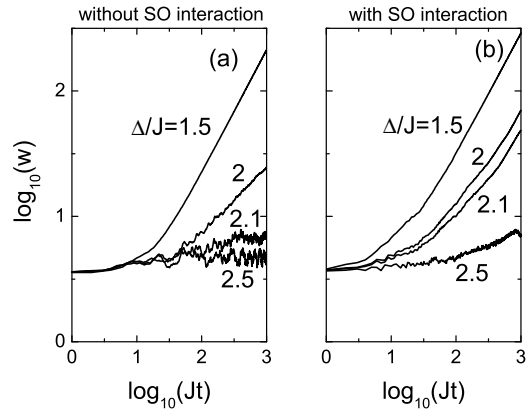}\caption{{}{\protect\footnotesize Time
evolution of the wavepacket width }$w\left(  t\right)  $%
{\protect\footnotesize . (a) without SO interaction; (b) in the presence of SO
interaction with }$\gamma=0.7$ {\protect\footnotesize and }$\Omega
/J=0.2${\protect\footnotesize .}}%
\label{fig_width}%
\end{figure}

\begin{figure*}[tbh]
\includegraphics[width=16cm]{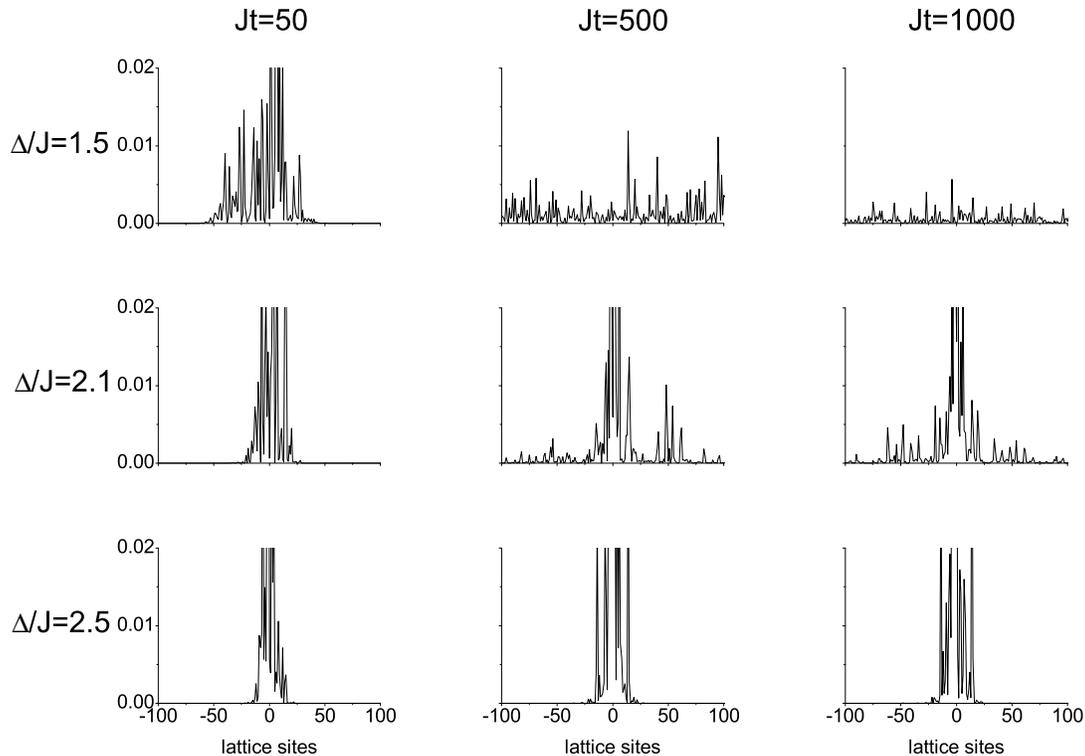}\caption{{\protect\footnotesize Time
evolution of wavepackets }$\left\vert \psi_{j,\uparrow}\right\vert ^{2}%
${\protect\footnotesize correspond to Fig. \ref{fig_width}(b) at specific
times.}}%
\label{fig_wavefunction}%
\end{figure*}

The above time-evolution properties are demonstrated in Fig. \ref{fig_width}%
(a). In Fig. \ref{fig_width}(b) we present the results in the presence of SO
interaction. One can find out that, for $\Delta/J=2.1$, which corresponds to
the critical phase shown in Fig. \ref{fig_phase_d-o}, the wavepacket still
subdiffuse with time evolution. In addition, the time evolution of wavepacket
at sample time are shown in Fig. \ref{fig_wavefunction}. At $\Delta/J=1.5$
correspond to the system in the extended phase, the wavepacket rapidly
diffuses and almost all the lattice sites become populated. While at
$\Delta/J=2.5$ for the localized phase of the system, there are no diffusion
because in this regime the initial Gaussian wavepacket can be decomposed into
the superposition of several single-particle localized eigenstates. For the
system in the critical phase at $\Delta/J=2.1$, the wavepacket diffusion is
accompanied with solitonlike structures in the centre and spreading sideband,
which reflects that extended and localized eigenstates coexist in the system.

Besides that, the nature of localized states can also be extracted from the
momentum distribution of the atom stationary states. This is because a more
localized atomic wavefunction corresponds to a wider momentum distribution via
Fourier transformation. The momentum distribution can be measured by turning
off the Raman lasers, releasing the atoms from the lattice and performing
time-of-flight imaging. Since our above discussions are for the dressed spin
states $\left\{  \left\vert \uparrow,k\right\rangle _{d},\left\vert
\downarrow,k\right\rangle _{d}\right\}  $, their momentum distribution can be
mapped from that of the bare spins $\left\vert \uparrow,k+k_{R}\right\rangle $
and $\left\vert \downarrow,k-k_{R}\right\rangle $ in the following manner%
\begin{align*}
\left\vert \uparrow,k\right\rangle _{d}  &  =\frac{1+i}{2}\left\vert
\uparrow,k+k_{R}\right\rangle +\frac{-1+i}{2}\left\vert \downarrow
,k-k_{R}\right\rangle ,\\
\left\vert \downarrow,k\right\rangle _{d}  &  =\frac{1+i}{2}\left\vert
\uparrow,k+k_{R}\right\rangle +\frac{1-i}{2}\left\vert \downarrow
,k-k_{R}\right\rangle .
\end{align*}

\section{Conclusion}

\label{sec_conclusion}

In conclusion, we have studied the system of a SO-coupled spin-1/2 particle
moving in a one-dimensional quasiperiodic potential. We mapped out the system
phase diagram in the tight-binding regime and accordingly discussed the
localization properties. In the absence of SO interaction the system can be
mapped into the AA model and self-dual if $\Delta/J=2$, SO interaction breaks
the duality and leads to the appearance of critical phase, in which the
extened and localized states coexist in the energy spectra. We also verified
the phase diagram via multifractal analysis of the wavefunctions and diffusion
dynamics of a initially localized Gaussian wavepacket. Experimental detection
of localization properties are discussed. We proposed an experimental
realization of the system using cold atomic gas trapped in a quasiperiodic
optical lattice potential and external laser fields. Since the two ingredients
of our proposed scheme, the quasiperiodic optical lattice potential
\cite{inguscio2008} and SO coupling
\cite{spielman2011,zhang2012,zwierlein2012}, had already been achieved for
cold atoms, it is expected that the localization properties discussed in this
work can be readily observed in experiment.

\begin{acknowledgments}
This work is supported by the National Basic Research Program of China (973
Program) under Grant No. 2011CB921604, the National Natural Science Foundation
of China under Grant Nos. 11234003, 11129402, 11004057 and 10828408, the "Chen
Guang" project supported by Shanghai Municipal Education Commission and
Shanghai Education Development Foundation under grant No. 10CG24, and
sponsored by Shanghai Rising-Star Program under Grant No. 12QA1401000. H.P. is
supported by the US NSF, the Welch Foundation (Grant No. C-1669), and the
DARPA OLE program.
\end{acknowledgments}

\end{document}